# Opto-Electronics in Large Array Gas Detector Systems

M. R. Dutta Majumdar, Debasish Das and Tapan K. Nayak
Variable Energy Cyclotron Centre,
1/AF, Bidhan Nagar, Kolkata - 700064, INDIA

## ABSTRACT

Large array gas detector systems are used in particle and nuclear physics experiments involving high-energy nucleon-nucleon and heavy-ion collisions. We have observed that in large array gas detector systems the momentary discharges inside the detector cells result in slowdown of the High Voltage conditioning and possible hindrances in signal processing. We have explored the opto-electronic devices like the opto-coupler, optical fibre and signal processing circuit, which provide successful monitoring procedures and preventive measures to overcome the challenges produced in such complex detector systems.

## 1. INTRODUCTION

Large array gas detector systems are widely used in particle and nuclear physics experiments to study and characterize the collision environments in ultra-relativistic nucleon-nucleon and heavy-ion reactions. One of the primary goals of such high-energy heavy-ion experiments is to study the formation and characteristics of the quark-gluon plasma (QGP), a state of matter, which may be created at sufficiently high-energy temperatures and densities. Detecting and understanding the QGP phase will allow us to understand the state of Universe in the moments after the Big Bang. Because of the density of produced particles from the collision environment in such reactions is quite high, the gas detectors should be highly segmented or granular to suit for such high multiplicity nuclear physics experiments.

Large number of charged particles and photons are produced in high-energy heavy-ion collisions. One of the probes of QGP formation is the measurement of photon multiplicity and spatial distribution of high-energy (~GeV range) photons on an event-by-event basis. The basic principle of photon measurement involves a pre-shower technique using a lead converter and measuring the produced shower particles to characterize the photons. A high-energy photon produces an electromagnetic shower on passing through the converter. These shower particles produce signals in the sensitive volume of the detector. The thickness of the converter is optimised such that the conversion probability of high-energy photons is high and transverse shower spread is small to minimize shower overlap in a high multiplicity environment. The Photon Multiplicity Detectors (PMD) have been successfully used in WA93 and WA98 experiments at CERN, Geneva[1,2] at the CERN Super Proton Synchrotron. The sensitive medium of PMD in these experiments was plastic scintillator pads, which were coupled with Wave Length Shifting (WLS) optical fibres. The signals were read through Charge Coupled Device (CCD) cameras with Image Intensifiers. It consisted of highly segmented detector placed behind a lead converter of suitable thickness.

In the ongoing International Collaborations at Relativistic Heavy Ion Collider (RHIC) at Brookhaven National Laboratory, USA and the future experiment at the Large Hadron Collider (LHC) at CERN, Geneva we have decided to shift the detector methodology from plastic scintillator to highly segmented gas detectors. The PMD[3] presently installed at the STAR experiment at RHIC is a large array gas detector system comprising of 82944 hexagonal honeycomb shaped cellular proportional counters of cross-section 1.0 square cm and gas depth of 0.8 cm. The gas mixture of $Ar+CO_2$ (70:30) is used as the sensitive medium.

It has been observed that in large array gas detector systems like PMD, momentary discharges inside the detector cells resulted in the tripping of High Voltage (HV) supply and also causing an occasional damage of signal readout processors. The initial inclusion of opto-electronic devices like the opto-coupler[4] provides an additional sensor to monitor such momentary discharges taking place in the gas detector system.

## 2. METHODOLOGY

**2.1 Description of Large Array Gas Detector Systems**

The PMD is a modular detector consisting of 12 gas tight enclosures called SuperModules in each planes and there are two planes. The detector is based on a proportional counter design using $Ar + CO_2$ gas mixture. This gas mixture is preferred because of its insensitivity to neutrons. A super module is made of 4 to 9 unit modules as shown in Fig 1. Each unit modules consist of an array of 24 x 24 hexagonal cells arranged in 9 groups. The copper honeycomb cathode is kept at the operating value of -1520V through a High Voltage filter box as shown in Fig 2 (a). The tungsten anode wire (of 20μm diameter) is at ground potential.

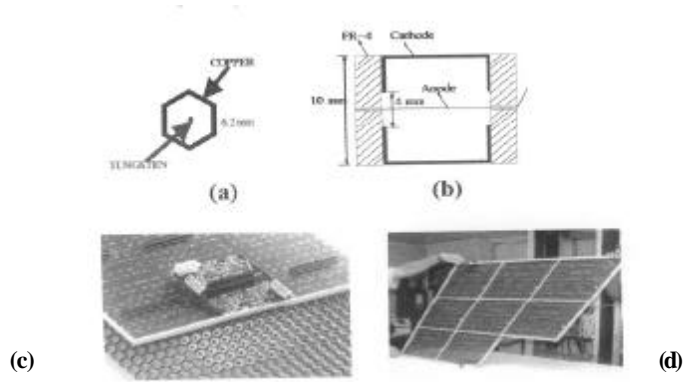

Fig. 1. (a) Unit cell schematic (b) Unit Cell cross-section showing dimensions and the cathode Extension. (c) Components of a unit module: Copper honeycomb, placed between two PCBs. The top PCB is seen with connectors and a FEE board. The cathode extension on the inside of the bottom PCB and the island separating the anode wire with the cathode is visible through the honeycomb. The photograph was taken with unassembled components. (d) An assembled Super Module with 8 Unit Modules. Divisions within a SuperModule denote Unit Modules

The Front-End Electronics (FEE) for processing the PMD signals is based on the use of 16-channel GASSIPLEX[5] chips developed at CERN, which provide analog multiplexed signals, and readout using the custom built ADC board. The chip has a peaking time of 1.2μs and the gain is 3.6 mv/fC over a linear dynamic range of 560fC. Considering the symmetry requirements of the detector hardware, the readout of the entire PMD has been divided into 48 chains. Each chain covers three unit modules and has 1728 channels.

The cells in the unit modules are arranged in 9 groups consisting of 8x8 cells connected to a 70-pin connector. A FEE[6] having four GASSIPLEX chips reads out this cluster of 64 cells. One such board is shown in Fig. 2(b). Keeping in view the geometrical considerations the FEE board is also made in rhombus shape. When all the boards are placed on the detector, they almost fully cover the detector area.

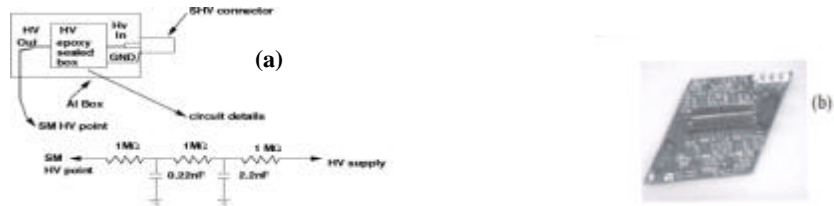

Fig 2 (a) High Voltage filter schematic (b) A FEE board with 4 Gassiplex chips

## 2.2 Principle of Opto-Electronic Devices

Opto-electronic devices are composed of several sub-systems like light emitters, light transporting medium and light sensors. Light emitters are essentially light emitting diodes (LED) normally emitting radiation in infrared or visible region. Light is normally transported through a thin dielectric medium or through an optical fibre to a light sensor. Light sensors are generally silicon diode or Photo Multiplier Tubes (PMTs). The monitoring current of an electrically isolated system is made to pass through light emitter and the generated light is transported to a light sensor to make an electrically isolated output. As the output signal transfer is due to optical coupling, the arrangement is called an opto-coupler. In this type of opto-electronic device the light emitter's light yield is required to be high and wavelength matched with the conversion (or quantum) efficiency of the light sensor. In case of silicon photo diode the wavelength match occurs in the infrared region. Alternately high gain photo multiplier tube (PMT) with bi-alkali type photo cathode needs Green emitting light source or such green LEDs to suit with the quantum efficiency of photo-cathode.

## 2.3 Application of Opto-electronics in Large Array Gas Detector

The SuperModules of the PMD, which consist of large array honeycomb cells enclosed in a gas tight chamber, have all been tested in the laboratory[7] using the following procedure:

- Test the gas leak in the SuperModules with a sniffer probe.

- Test of High Voltage (HV) stability by measuring leakage current and monitoring discharge.

- Test of channel-by-channel pedestal and its variation at the operating voltage using FEE and a PC-based Data Acquisition System[8].

- Test to observe the pulse signal with strong radioactive source.

Testing of HV stability was started after about 24 to 48 hours depending on the size of the SuperModule. For the test with high voltage, the 9 readout zones in a Unit Module were shorted with shorting connectors and the leakage currents were monitored up to -1600V. This is much higher than the operating voltage of –1450V. The set trip current limit was 2μA for a SuperModule. Groups of 9 connectors were sequentially added till the entire SM was covered with shorting connectors. It was observed that a group (64 cells) shorted with shorting connectors showed occasional tripping in presence of HV showing high current. The current was ≥ 5 μA both at lower (50V-400V) and higher voltages, indicating one or more bad channels in that group.

The proper monitoring of such occasional discharges was done with an opto-coupler circuit and hence providing a scope of such study in large array gas detector systems. The opto-coupler device was studied as follows:

- Online High Voltage monitoring of Super-Modules during initial testing, where the designed opto-coupler device was connected in series with a High Voltage box as shown in Fig 3.

- Testing of SuperModules require substantial time to trouble shoot problems at various levels starting from Unit Module to readout level and cell level. The opto-coupler (sense circuit) along with the Dual Inline Package (DIP) switch board and DVM was used to monitor locally the leakage current of a readout group (64 cells) in a Unit Module. The DIPswitch schematic is shown in Fig 4.

- The study involved identifying spark rates, sparking duration, nature of spark magnitude on the leakage current of the detector and its possible variations.

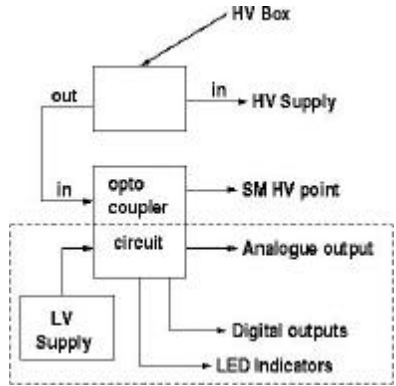
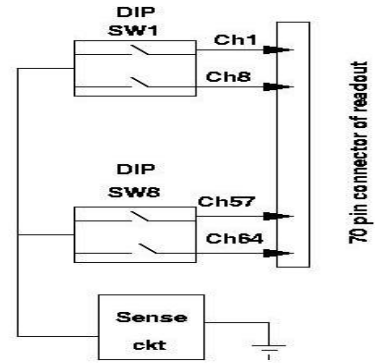

Fig 3. High Voltage layout with opto-coupler        Fig 4. DIP Switch schematic

### 2.4 Configuration of Opto-electronic device and fabrication

Momentary discharges in gas detector array is sensed by conventional opto-electronic Integrated Circuit but the light output of the sensing LED vary due to low level current, typically from a fraction of micro ampere to milliampere range. The non-linear behaviour observed below the operating range of existing conventional market components impose a challenge during its proper implementation. So our initial endeavour was with some discrete components and devices available to us. These were not vendor specific, but some important typical specification of light emitters and sensors are listed in Table 1 (a) and Table 1 (b).

Table 1 (a)

| Light Emitter | Operating Voltage | Emission Peak Wavelength in nm | Dimension (Diameter) | Emission Output | Response Time | Remarks |
|---|---|---|---|---|---|---|
| LED IR | 5 volt | 900 | 5 mm | ~1% | 0.5 – 1.0 µs | Plastic Package |
| LED Green | 5 volt | 500 | 5 mm | ~1% | 0.20 µs | Plastic Package |

Table 1 (b)

| Light Sensor | Dark Current | Operating Voltage in Volt D.C. | Quantum Efficiency – 880nm (IR) | Quantum Efficiency - 500nm (Green) | Active Area | Gain | Response time | Remarks |
|---|---|---|---|---|---|---|---|---|
| Photo Diode | 2 nA | 5 to 25 | 0.8 to 0.9 | 0.4 to 0.5 | 2.5 by 2.5 mm | No gain | 0.5 µs | Silicon in Plastic Package |
| PMT | 250 nA anode | -1100 | - | 0.15 | 13 - 19 mm diameter | $3 \times 10^6$ | 20 ns | Bi-alkali Cathode |

PMT is based on photo-electron generation from photo-cathode and multiple electron emission resulting very high gain (typically $3 \times 10^6$). Optical coupling medium like thin dielectric medium or optical fibre should have good transmission and minimum absorption to the emitted radiation. In case of thin dielectric medium coupling between light emitter and sensor is good. But such combination is hindered by electro-magnetic interference at the signal processing electronic device. Where as in optical fibre coupling arrangement, coupling is attenuated due to fibre length and numerical aperture. This combination in turn provides better signal processing and is free from electromagnetic interference. To achieve better performance using optical fibre one needs good transmission characteristics of emitted radiation, light gathering capability (in terms of high numerical aperture) and optical coupling at light source-sensor end. Multimode high numerical aperture plastic optical fibres are useful for present application.

We tested four basic configurations as shown in Fig 5. In fig 5(a) opto-coupler is made from truncated transparent infrared LED and photo diode. They are sand-witched after proper alignment between transparent thin plastic sheet with optical glue having optically coupling but electrical isolation up to the 2500 Volts. The described opto-electronic device is kept in light tight and electrically insulated enclosure. This configuration widely used in our work. Before sandwiching with the photodiode, the truncated LED was carefully machined and polished to make good optical coupling.

In the four basic configurations we use infrared LED along with photo diode and transparent green LED with high gain PMT. To enhance the compactness of the opto-electronic devices and avoid electromagnetic interference, plastic fibres are coupled with IR LED's to transmit the signal away from the gas detector as shown in Fig 5(b). Suitable cavity was made on LED to insert the 1.0 mm diameter plastic optical fibre made of Polystyrin core and Poly Metha Metha Acrylate (PMMA) clad. After insertion the portion is glued with optical glue. The numerical aperture of plastic fibre is 0.57. Other end is glued to fibre end-coupler-ferrule where the photo diode also connected under light tight condition.

In Fig 5(c) green LED is coupled to the cathode surface of PMT, which replaces the photo-diode in the previous configuration as described in Fig 5(a). Similarly the photo-diode arrangement described in Fig 5(b) is replaced with PMT in Fig 5(d). These two configurations are considered to observe low level cumulative signal pulses. It is possible to detect very low optical response with PMT and hence such configurations assumes paramount importance.

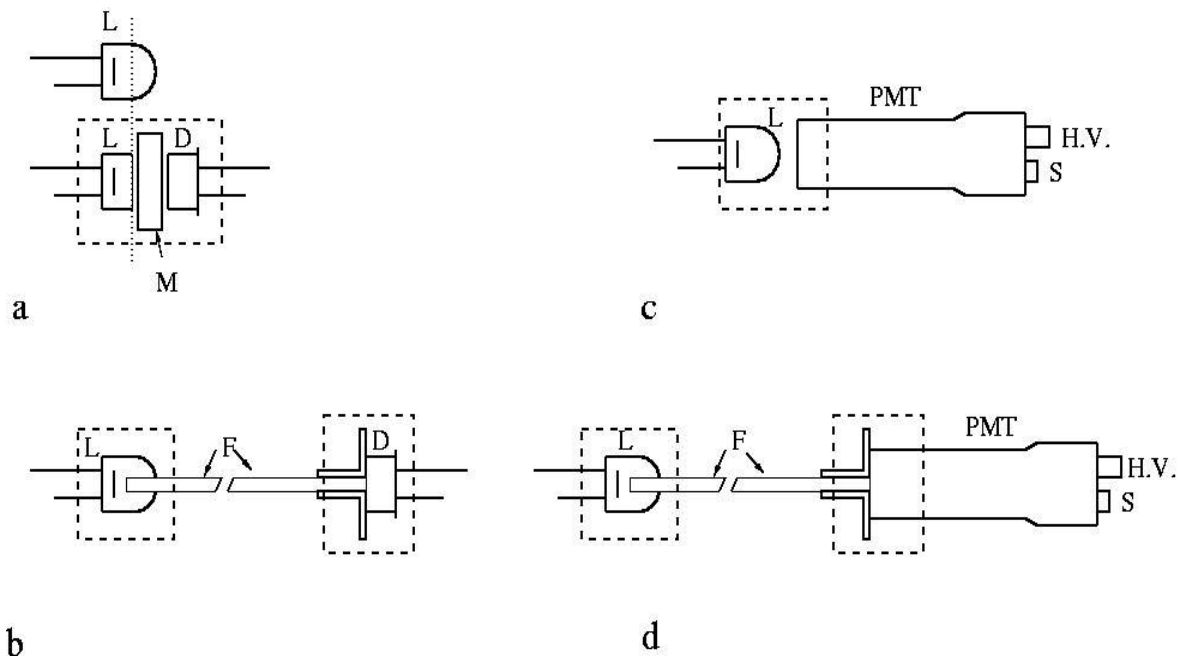

Fig 5. Configuration of Opto-Electronic devices

## 2.5 Design of opto-coupler circuit and testing

The designed circuit of opto-coupler signal processing arrangement is shown in Fig 6.In this circuit the photo-diode is amplified by Darlington pair. Emitter of the second transistor is coupled to a high value resistance (R1) to monitor leakage current of the SuperModule being tested and another resistance (R2) of lower value through the diode (D2) to monitor the pulse signals of varying nature. Pulse output is connected to a comparator and the output is high if it crossed the desired threshold level. High output logic of comparator triggers both fast and slow monostable circuits. Finally the output of comparator after monostable circuits and suitable gating generated logic pulses due to (i) a spike or cumulative effect of expected large number of short duration ($\leq$ 10µs) signals, (ii) small sparks (20 to 75µs typical) and (iii) long duration sparks ($\geq$ 75µs) or large DC level (caused due to shorting of anode wires). All three logic outputs gave visual indication by LEDs of three different colours. Logic of slower output is monitored using the counter and timer as shown in the schematic diagram. Functionality test of the fabricated circuit is done with varying current source (1 µA to 1 mA) in D.C. mode and in pulse mode with a suitable pulse generator having time duration from 2-300µs.

Finally circuit was tested with large array gas detector system and necessary data were taken to understand and infer its full functionality in such an experimental environment. We have found that the long duration discharges (typically $\geq$ 75µs) are mainly responsible for the damages in signal processing FEE boards. Henceforth our observation is focussed on such long duration discharges in greater detail. Here the circuit details described was for the configuration (a) of Fig 5.We have also tested with other combinations as described in section 2.4 and explained by schematic diagrams (b), (c) and (d) of Fig 5.

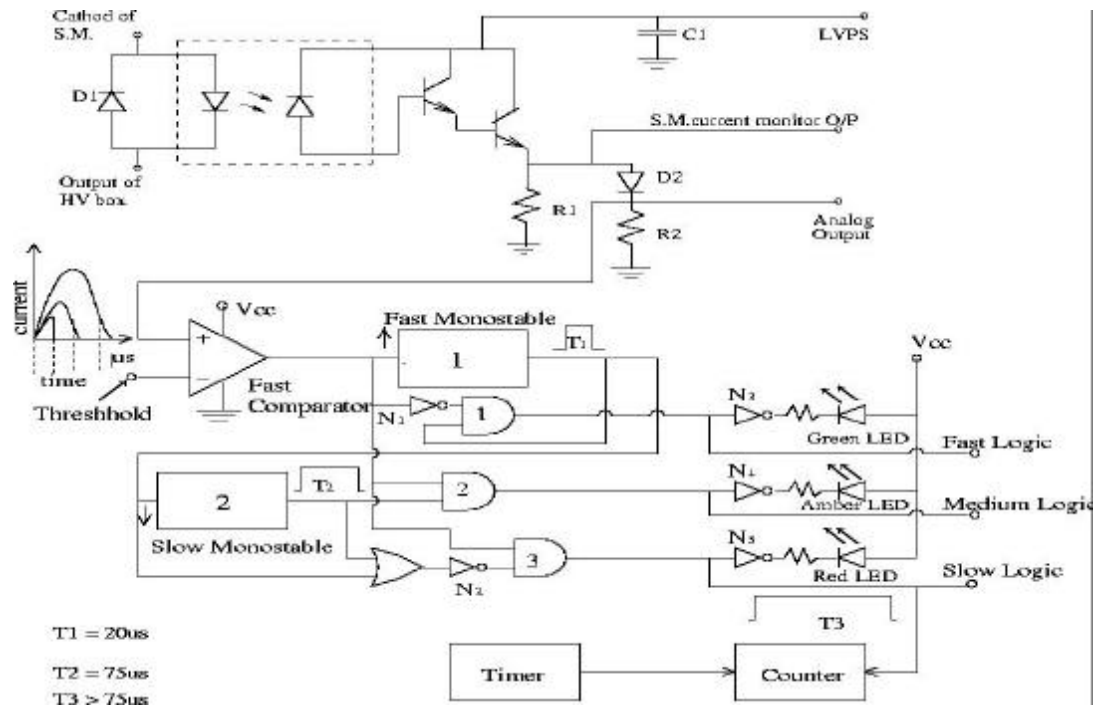

Fig 6. Schematic of Opto-Coupler signal processing circuit

## 3. DATA AND RESULTS

We connected the opto-coupler device with PMD and monitored the discharges (or sparks) to estimate the collective behaviour of such a large array gas detector system in presence of High Voltage. We have monitored the frequency of momentary discharges (or sparks) of long duration (typically $\geq 75\mu s$) in Unit Module level and readout level (comprising of 64 cells) with time in presence of –1600V as shown in the graphical representations of Fig 7.

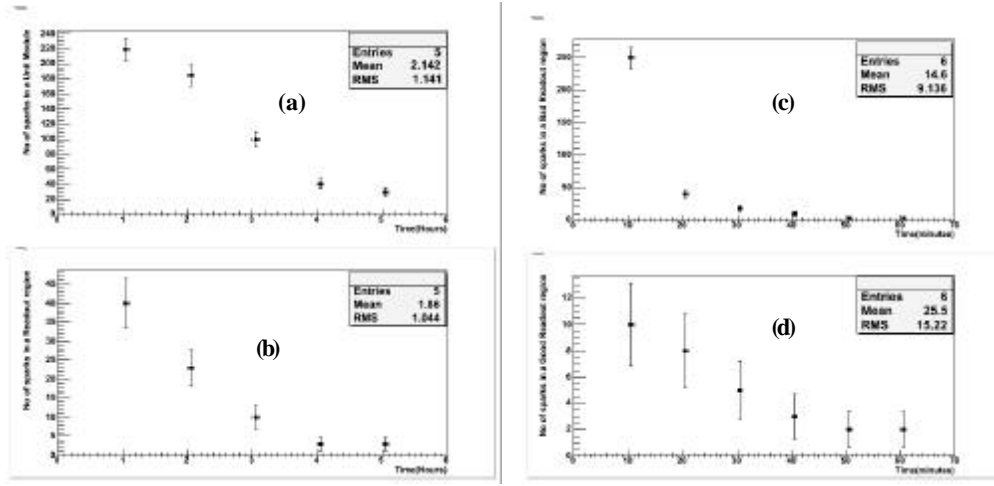

Fig 7. (a) The graphical representation of the frequency of sparks in one Unit Module (comprising of 576 cells) with time in presence of -1600V. We observe that the frequency of sparks reduce with time and hence indicating proper seasoning with HV. The time is measured in hours to ensure the stability of the system with HV. (b) Frequency of sparks in a typical readout region (comprising of 64 cells) with time. It also confirms proper seasoning. (c) The presence of a faulty cell in one bad readout region causing large number of sparks in a short time (measured in minutes). The opto-coupler ensures proper identification of that faulty cell with DIP switch as conveyed in section 2.3 . The bad cell is removed and the readout region after removal behaves like a normal good readout region showing proper seasoning with time. (d) A good seasoned readout region. The error bars are purely statistical.

The 4 configuration details of opto-electronic device used for our present study are summarized in tabular form in Table 2.

Table 2

| Configuration | Operating Current range | Response Time | Application | Remarks |
|---|---|---|---|---|
| LED (IR) + Photo diode | a) 2 - 10 µA b) 25 - 500 µA | 5 µs | a) Leakage current b) Monitoring discharges | Widely used |
| LED (IR) + Fibre + Photo diode | a) 2 - 10µ A b) 25 - 500 µA | 5 µs + delay of fibre | a) Leakage current b) Monitoring discharges | Functionally Good, needs some Optimisation |
| LED (green) + PMT | a) 2 - 10 µA b) 1 - 50 µA | 30 ns | a) Low level leakage current b) Signal behaviour | Needs more controlled study |
| LED (green) + Fibre + PMT | a) 2 - 10 µA b) 1 - 50 µA | 30 ns + delay of fibre | a) Low level leakage current b) Signal behaviour | Needs more controlled study |

## 4. CONCLUSIONS

Opto-electronic devices are of valuable use especially for offline monitoring of the faulty cells in such large array gas detector systems like PMD. Present experience from the testing of opto-electronic device with plastic optical fibre will provide useful experience to implement for the online monitoring during high-energy heavy-ion physics experiments in future. Further study initiated with PMT may help in estimating large array gas detector occupancy (number of detector cells active compared to total cells of the system) in a large multiplicity experimental scenario. After some more study we wish to make final modular devices so that they can be incorporated during the up gradation plan of STAR detector and also in other future experiments like LHC.

## ACKNOWLEDGEMENT

The authors are grateful to Dr. Bikash Sinha, Director, Variable Energy Cyclotron Centre (VECC), Dept of Atomic Energy (DAE), Government of INDIA, for his encouragement and support. Valuable advice of Dr Y. P. Viyogi, spokesperson of PMD team and Project Manager, VECC, is gratefully acknowledged. Discussion with PMD collaboration colleagues on various aspects of detector and present opto-electronics work is also acknowledged.